\documentclass[12pt,preprint]{aastex}

\slugcomment{Accepted to appear in Icarus}

\shorttitle{Light Curve for the KBO 2000 $EB_{173}$}
\shortauthors{Schaefer \& Rabinowitz}

\begin{document}

\title{Photometric Light Curve for the Kuiper Belt Object 2000 $EB_{173}$ on 78 Nights}

\author{Bradley E. Schaefer}
\affil{Department of Astronomy, University of Texas,
    Austin, TX 78712}

\author{David. L. Rabinowitz}
\affil{Yale University, New Haven, CT 02511}

\begin{abstract}
Kuiper Belt Objects (KBOs) are generally very faint and cannot in practice 
be monitored with a well-sampled long-term light curve; so our discovery of 
the bright KBO 2000 $EB_{173}$ offers an excellent opportunity for synoptic 
studies.  We present a well-sampled photometric time series (77 R magnitudes 
and 29 V magnitudes on 78 nights) over a 225-day time span centered on the 
2001 opposition.  The light curve (corrected to the year 2001 opposition 
distance) varies from 19.11 to 19.39 mag with a single peak that is smooth, 
time symmetric, and coincident with opposition.  All variations in the light 
curve are consistent with a linear opposition surge ($R_{OPP} = 19.083 + 0.125 
\alpha$, where $\alpha$ is the solar phase angle), while any rotational 
modulation must have a peak-to-peak amplitude of less than 0.097 mag.  
This is the first measured opposition surge for any KBO (other than Pluto).  
The \vr ~ color is $0.63 \pm 0.02$, with no apparent variation with phase at the 
few percent level. With $R=19.11$ at opposition, 2000 $EB_{173}$ remains the 
brightest known KBO and a prime target for future photometric and spectroscopic 
studies.
\end{abstract}


\section{Introduction}

Most searches for Kuiper Belt Objects (KBOs) scan over small areas of the sky 
to deep limits, with little chance of discovering the rare-bright objects.  
Only in the last two years have a number of studies searched wide areas 
\citep{fer01,lar01,she00,jlt98}. As part of the QUEST KBO Survey ($66.8 
deg^2$ to $R=20.1$), we discovered the very bright KBO 2000 $EB_{173}$ 
\citep{fer01}.  This body is one of the largest known KBOs, and the 
brightest after Pluto.

As a bright source, 2000 $EB_{173}$ can provide the best signal-to-noise 
ratio for a variety of physical observations, including visible and 
infrared spectroscopy.  Our own discovery and initial follow-up photometry 
\citep{fer01} established the red color of 2000 $EB_{173}$, typical of many 
KBOs and attributed to the presence on the surface of organic solids 
\citep{cru98}. \citet{bbk00} give a spectrum from 1.4-2.4 {$\mu$}m, 
\citet{lod01} present a spectrum from 0.9-2.4 {$\mu$}m, and \citet{jel01} 
give a reflectance spectrum from 1.0-2.4 {$\mu$}m.  The infrared spectrum 
of 2000 $EB_{173}$ is generally featureless although Licandro et al. note 
strong absorption in the K band.  This might be related to highly carbonized 
materials with a deficiency of hydrogen bonds.

Here, we report the first long-term photometric light curve for 2000 $EB_{173}$ 
measured on 78 nights from December 2000 to July 2001 with the Yale 1-meter 
telescope at Cerro Tololo Observatory. The data reveal a prominent opposition 
surge, which is a significant increase in brightness within a few degrees of 
zero phase. The effect is commonly observed for the moon, asteroids, and 
satellites of planets in the outer solar system \citep{geh56}, and has been 
attributed both to the sudden disappearance of shadows cast by regolith 
particles and to coherent backscattering by particle aggregates 
\citep{shh01,hap93}. For the high-albedo icy satellites in the outer 
solar system, laboratory investigations suggest that coherent backscattering 
may play a dominant role \citep{nel98}. For the well-studied asteroids in the 
main belt, the shape of the phase curve is known to correlate with spectral 
type and albedo \citep{bes00,hev89,bol79}. In this paper we compare the phase 
curve we observe for 2000 $EB_{173}$ with model predictions and with other 
solar system bodies to place constraints on the likely composition and surface 
properties.

\section{Observations}

	We made all our photometric observations of 2000 $EB_{173}$ with the 
Yale 1-m telescope at Cerro Tololo operated by the YALO consortium 
\citep{mdb98,bai99}.  As observations with this telescope are queue-scheduled 
by a resident operator, this mode is well suited for secular observations. 
This capability and the accessibility of 2000 $EB_{173}$ at our modest aperture 
provided an opportunity to obtain the first long-term light curve of a KBO (other 
than Pluto) with fine time resolution.

	All our data are from the ANDICAM CCD camera \citep{mdb98}.  The array has 
$2048 \times 2048$ pixels, each $0.29 \arcsec$ square.  The read noise is 11 $e^-$ 
on average, and we operated at a gain of $\sim 3.6 e^- ADU^{-1}$.  All our exposures 
were of 300 second duration.  Our typical seeing condition resulted in a FWHM of 
$\sim 1.5 \arcsec$ (5 pixels). With a maximum apparent motion for 2000 $EB_{173}$ 
of $99 \arcsec ~ day^{-1}$, the trailing was less than 1.2 pixel.  A detailed 
calculation shows that this trailing results in a negligible systematic 
photometric error (under 0.001 mag for our 5 pixel radius photometry aperture).

	Our observations of 2000 $EB_{173}$ cover 225 days from JD 2451890 
(11 December 2000) to JD 2452115 (24 July 2001).  We discarded 29 images 
because the image fell on a bad column, a cosmic ray hit, or a background 
star.  In all, we obtained 214 useable images on 78 separate nights.  Of 
these, 185 are in the Johnson R-band with a fairly even distribution across 
our observing interval, while 29 are in the Johnson V-band with all but one 
of these taken before JD 2451942 (1 February 2001).  By averaging repeated 
measurements taken within 15 minutes on the same night, and weighting 
inversely by the estimated variance, we obtain 77 R-band and 28 V-band 
magnitudes. They are listed in Table I with 1-sigma errors and are plotted 
in Figure 1 versus observation date. For each observation Table I also gives 
the heliocentric Julian Date of the middle of the exposure(s) and the solar 
phase angle $\alpha$, derived using the JPL/Horizons program at 
http://ssd.jpl.nasa.gov/.  

Our method for measuring the magnitudes and determining their uncertainties 
follows the standard techniques in CCD photometry. We used IRAF programs 
(ZEROCOMBINE, FLATCOMBINE, CCDPROC) to determine mean bias and flat fields 
(from five bias frames and domes flats taken at the start of each observing 
night) and to make bias and flat field corrections for each image. We used 
IRAF program PHOT to measure the flux of objects within apertures of radii 
10 pixels for absolute photometry and within radii of 5 pixels for relative 
photometry.  For each object we determined the sky background as the mode of 
the flux distribution within an annulus of inner and outer diameters 15 and 30 
pixels.

Our calibration procedure consisted of three steps.  In the first, we analyzed 
images of Landolt photometric standard stars \citep{lan92} taken on eight 
photometric nights.  Typically 29 stars were observed on each night with 
airmasses ranging from 1.15 to 1.7.  For each night we fit Landolt's 
catalogued magnitudes for these stars to a linear function of the instrumental 
magnitudes, the \vr ~ colors were measured, and corrected for their airmass at 
the time of observation.  The standard deviations for these fits range from 
0.010 to 0.030 mag (with a median of 0.015 mag), this being a measure of our 
systematic uncertainties.

Our second step was to determine the R- and V-band magnitudes of many 
comparison stars near to the path of 2000 $EB_{173}$.  We chose 151 stars 
such that every image of 2000 $EB_{173}$ included nine comparison stars on 
average. Their magnitudes range from 14 to 18, yielding images with large 
signals but not saturated.  The stars appear in 19 fields, each of which 
we recorded with 180-second exposures in both the R and V bands on the same 
nights that we observed the Landolt stars.  Using our fits to the Landolt 
magnitudes, we determined the V and R magnitudes of the comparison stars 
given their airmass and instrumental V and R magnitudes. The resulting 
comparison star magnitudes are accurate to approximately 0.01 mag.

Our third calibration step was to determine the R and V magnitudes of 2000 
$EB_{173}$ using aperture corrections, thus minimizing the contribution of 
the sky background to the noise in each measurement.  For each usable image 
of the object, we first measured its instrumental magnitude using a 5-pixel 
radius aperture. We then converted this result to a measurement within a 
10-pixel radius aperture by adding the median difference between the 5- and 
10-pixel instrumental magnitudes for nearby bright comparison stars. Finally, 
we determine the R-band or V-band magnitude for 2000 $EB_{173}$ from its 
corresponding 10-pixel instrumental magnitude by adding the median difference 
between calibrated and instrumental magnitudes for the comparison stars 
appearing in the same image. 

The dominant uncertainty in our resulting measurements for 2000 $EB_{173}$ 
is the uncertainty from the flux measurements within the 5-pixel apertures 
which is typically 0.04 to 0.08 mag.  On most of our nights, the R-band 
images were taken as pairs (from JD 2451890 to JD 2451942) or as triples 
(from JD 2451954 to JD 2452115) in quick succession.  The errors after 
combining the resulting flux measurements as weighted averages, which 
generally range from 0.02 to 0.05 mag., as shown in Table I.

	We have checked our determinations of these errors in two ways.  
First, we have many multiple measures of 2000 $EB_{173}$ that are nearly 
simultaneous. Their differences (in units of the combined one-sigma uncertainty) 
have an RMS scatter of 0.98, implying that our individual error bars are accurate.  
Second, we have produced magnitudes for eight check stars (with magnitudes close 
to that of 2000 $EB_{173}$) using the same procedures we used for 2000 $EB_{173}$.  
Fitting constant light curves to each of the stars we find the total chi-square 
to be 172 for 164 degrees of freedom.  The reduced chi-square is near unity, so 
we conclude that our method for deriving the uncertainties is valid.

	For the purposes of studying the long-term variability of 2000 $EB_{173}$, 
we have calculated the 5-day average magnitudes and errors shown in Table II and 
plotted in Figure 2 as a function of solar phase angle $\alpha$. We computed the magnitudes 
from the R-band data in Table I for each successive five-day intervals starting at 
JD 2451890.  For each average magnitude, Table II also shows the average Julian 
date and solar phase angle for each 5-day average.  We corrected the observed 
magnitudes (R) for the changing geocentric distance of 2000 $EB_{173}$ ($\Delta$) 
using $R_{OPP} = 5 \log_{10}(\Delta /28.76)$ where $R_{OPP}$ is the R-band magnitude 
corrected to the opposition distance.  The geocentric distance of 2000 $EB_{173}$ 
changed from 30.31 AU at the start of our observations to 28.76 AU at opposition, 
for a correction of up to 0.11 mag.  The heliocentric distance of 2000 $EB_{173}$ 
varies from 29.80 AU to 29.70 AU over our interval of observation, resulting in a 
totally negligible correction of 0.007 mag.

\section{Analysis}

	The light curve for 2000 $EB_{173}$ shown in Figure 1 has a prominent peak 
at JD $2452010 \pm 10$ that is broad, symmetrical, and that coincides with the date 
of opposition on JD 2452010 (with $\alpha=0.28 \degr$). There are also two minima at 
JD $2451915 \pm 10$ and $2452100 \pm 10$, each coinciding with the respective times 
of maximum phase on JD 2451922 and 2452102 (with $\alpha$ equal $1.89 \degr$ and 
$1.96 \degr$, respectively).  These coincidences clearly suggest that the brightness 
variations we observe for 2000 $EB_{173}$ are phase dependent, and the result of a 
significant opposition surge.

	The phase curve for 2000 $EB_{173}$ plotted in Figure 2 shows that this 
opposition surge is linear over the observable range of phase. A least-squares 
fit yields $R_{OPP} = 19.083 + 0.125 \alpha$.  The one-sigma uncertainties in 
the constant and phase coefficient are $\pm 0.005$ mag and $\pm 0.009 ~ mag 
~ deg^{-1}$.  Because the reduced chi-square for this linear fit is close to 
unity ($\chi ^2 = 1.10$), there is no justification for fitting a more 
complicated model for the phase function (e.g., a curve or a broken line).  
Correcting the unbinned data from Table I to opposition distance and again 
fitting a linear phase function also yields a reduced chi-square close to 
unity (1.31).  At this finer resolution the RMS scatter about a linear fit 
is 0.059 mag, while the average uncertainty of the input magnitudes is 0.048 
mag. 

	In terms of the parameters introduced by \citep{hap93} to characterize 
the phase functions of various solar system bodies, our observations of 2000 
$EB_{173}$ provide significant constraints. We have fit the phase curve with 
Hapke's integral phase function for the opposition surge component, which to 
first order in $\alpha$ is proportional to $[1.0+B(\alpha)]p(\alpha)-1.0$, where 
$p(\alpha)$ is the phase function for a single particle and $B(\alpha) \cong 
B(0)[1.0+(1/h) \tan (\alpha /2)]-1$. Here $B(\alpha)$ describes an opposition 
surge of  amplitude $B(0)$ and angular half-width $h/2$ radians. Values for 
$B(0)<1$ are expected for surges resulting solely from shadow-hiding, whereas 
values exceeding unity may occur due to coherent backscattering.  For our small 
angular range of $\alpha$, $p(\alpha)$ is very close to one, so then the 
opposition surge component will be $B(\alpha)$ and the total reflected 
intensity will scale as $1+B(\alpha)$.  In magnitudes, $R=R_0 -2.5 
\log_{10}\{[1+B(\alpha)]/[1+B(0)]\}$, where $R_0$ is the magnitude at 
zero phase. Given the narrow phase range of our observations, we are not 
able to constrain either the albedo or the surface roughness, which enter 
into Hapke's expressions at higher orders in $\alpha$.

	Our best fit yields $h = 0.050$ radians ($2.9 \degr$), $B(0)=4.7$, 
and $R_0=19.07$; with a chi-square of 37.6 for 35 degrees-of-freedom. From 
a Monte Carlo analysis of our observations, we derive a one-sigma 
uncertainties $\pm 0.015$ rad for $h$, -3 and +2 for $B(0)$, and  $\pm 0.02$ 
mag for $R_0$.

We have many virtually simultaneous V-band and R-band measurements, so we can 
look for variations in the \vr ~ color.  The average \vr ~ color is $0.63 \pm 0.03$ 
from JD 2451893 to JD 2451923 (phase $1.678 \degr - 1.893 \degr$), $0.63 \pm 
0.02$ from JD 2451927 to JD 2451942 (phase $1.883 \degr - 1.767 \degr$), and 
$0.62 \pm 0.07$ for JD 2452030 (phase $0.745 \degr$).  Thus we find no evidence 
for any fast or slow variations.  The final \vr ~ color for all nights combined as 
a weighted average is $0.63 \pm 0.02$ mag.  This agrees with $\vr =0.57 \pm 0.05$ 
determined by Ferrin et al. (2001) from earlier observations.

	Because the \vr ~ color for 2000 $EB_{173}$ appears constant, our 
observations alone reveal no significant difference between the V-band 
and R-band phase curves. Including the \vr ~ observations presented by 
\citet{fer01} and \citet{jel01} we can refine this constraint.  In 
Table 3, we have collected all available \vr ~ measures for 2000 
$EB_{173}$.  Our chi-square fit to these data yields a \vr ~ phase 
curve with slope $0.03 \pm 0.04 ~ mag ~ deg^{-1}$.  This slope differs 
by less than one sigma from zero, so that again we find no significant  
color dependence for the opposition surge. Adding this marginally 
non-zero coefficient to our R-band phase coefficient of $0.125 
\pm 0.009 ~ mag ~ deg^{-1}$, we obtain a V-band phase coefficient of 
$0.155 \pm 0.041 ~ mag ~ deg^{-1}$.

\section{Implications}

One of the primary goals of our long-term study of 2000 $EB_{173}$ is 
to determine its rotation period.  However, we see no significant 
photometric modulations other than the ordinary phase function with 
its opposition surge.  Quantitatively, this is shown by the fit of 
nightly R magnitudes (see Table 1) having a reduced chi-square near 
unity.  However, it is possible that 2000 $EB_{173}$ has a very low 
amplitude modulation, and this might be shown by the nightly scatter 
about the phase function (0.059 mag) being slightly above the average 
one-sigma error bar (0.048 mag).  If this difference is ascribed to 
rotational modulation, then the peak-to-peak amplitude cannot be 
larger than 0.097 mag.  The error bars are too large to have any 
real hope of finding a significant periodicity with only 77 data 
points spread out over 225 nights.  

The lack of a rotational modulation is not surprising, as by our 
count from the literature 
\citep[for example]{far01a,far01b,gut01,dav98,shj02}, roughly 32 
KBOs and Centaurs have been checked for variability and this has 
yielded only 14 known rotational periods.  A flat rotational light 
curve could be due to 2000 $EB_{173}$ being viewed pole-on or due 
to its being nearly round and without large regional albedo differences.

	The absence of any significant rotational variability and 
the bright appearance of 2000 $EB_{173}$ relative to other KBOs has 
allowed us to measure the object's phase curve, the first for any KBO 
other than Pluto \citep{ras01}. In Figure 2,  we compare the phase 
curve to planet Pluto (the largest known KBO), outer solar-system 
satellites Rhea, Nereid, and Titania, and also to the dark main-belt 
asteroid (24) Themis.  These examples have been chosen because they 
span the range of small-angle phase curves that have been well measured 
for bodies that might have surfaces similar to 2000 $EB_{173}$. Nereid 
is particularly interesting because it might be a captured KBO 
\citep{scs95,scs00} and because it also exhibits a prominent opposition 
surge \citep{sct01}. 

It is apparent that 2000 $EB_{173}$ has an opposition surge with a slope 
intermediate in amplitude among the comparisons in Figure 2.  (This is 
also confirmed by comparing $V_{OPP}(0 \degr)-V_{OPP}(2 \degr)=0.31$ 
with the tabulated values in Table II of \citet{sct01}.) The surge is 
much steeper than for Pluto, Rhea, and most main-belt asteroids, but 
not as steep as the surges observed for Nereid and Titania. Unlike 
these two satellites which have narrow opposition surges, 2000 
$EB_{173}$ also has a wide surge comparable to Rhea and (24) 
Themis. Strong and narrow opposition surges have also been measured 
for particulate materials of high albedo, for which the cause is 
attributed to coherent backscatter \citep{nel98,nel00}. It is thus 
possible that the influence of coherent backscatter is modest for 
2000 $EB_{173}$ in comparison with Nereid and Titania. This might 
also indicate a relatively low albedo for 2000 $EB_{173}$. However, 
because the high-albedo surfaces of Pluto and Rhea (0.6) have even 
weaker opposition surges, this conclusion can not be drawn. Owing 
to the intermediate slope of the opposition surge in 2000 $EB_{173}$, 
the comparison with Nereid does not allow any confident conclusion 
concerning the possibility that Nereid is a captured KBO.  

	Despite the relatively broad opposition surge for 2000 
$EB_{173}$, the large amplitude that we derive ($B(0)>1.7$ with 
84\% confidence) shows that coherent backscattering is likely an 
important influence. This can be tested by looking 
for a wavelength dependence to the width of the 
opposition surge, $h$.  For shadow-hiding, there is no 
wavelength dependence. For coherent-backscattering, $h$ 
scale linearly with wavelength of observation \citep{hap93}.  
Our R-band observations yield $h=0.050$ rad. For the V-band, 
we should therefore  expect $h= 0.050 \times (5500 \AA /7000 \AA)=0.039$ 
rad if coherent-backscattering is the dominant cause, and 
$h = 0.050$ rad if shadow-hiding is dominant..  Unfortunately, 
our observations  are not sufficiently precise to measure this 
small difference. However, in the near future, a simple measure 
of the $B-I$ or $B-R$ color at opposition can be compared with 
the $B-I=1.97 \pm 0.13$ mag and $B-R=1.59 \pm 0.13$ mag reported 
for a phase of $1.73 \degr$ \citep{fer01} or $B-I=2.17 \pm 0.05$ 
mag and $B-R=1.58 \pm 0.04$ mag reported for a phase of $1.93 
\degr$ \citep{jel01}.  This comparison could provide a decisive 
test between shadow hiding and coherent backscattering.

\section{Conclusions}

We measured 77 R-band and 28 V-band magnitudes of 2000 
$EB_{173}$ on 78 nights over a 225 day interval centered on 
its 2001 opposition.  We find that the light curve varies from 
$R_{OPP}=19.11$ to $19.39$, with all the changes fully described 
by a linear phase function with $R_{OPP}=19.083+0.125 \alpha$. 
This is the first opposition surge measured for any KBO (other 
than Pluto).  The amplitude of any rotational modulation is 
small, with a peak-to-peak amplitude of less than 0.097 mag.  
The \vr ~ color is constant at $0.63 \pm 0.02$ mag.  At 
opposition (with $\alpha=0.28 \degr$), 2000 $EB_{173}$ is the 
brightest known KBO with $R=19.11$ mag.

	We are able to fit the R-band phase curve assuming 
a Hapke-type opposition  surge with width  $h=0.050$ and 
amplitude $B(0)>1.7$. This is relatively broad compared to 
Nereid, and does not offer confirmation that Nereid could be 
a captured KBO. The large amplitude of the surge, however, 
indicates the coherent backscattering may be important.  Our 
measurements of the phase curve in the V-band are not accurate 
enough to distinguish between shadow hiding and coherent 
backscattering.  Future measures of the $B-I$ or $B-R$ color 
near opposition could determine the physical cause of the 
opposition surge.

\clearpage

\begin{deluxetable}{cccc}
\tabletypesize{\scriptsize}
\tablecaption{Observed Brightness of 2000 $EB_{173}$ in 2000/2001. \label{tbl-1}}
\tablewidth{0pt}
\tablehead{
\colhead{$\langle HJD \rangle$} & \colhead{$\alpha$}   & \colhead{$R$}   &  \colhead{$V$}
}
\startdata
2451890.85	 & 	1.632	 &  	19.25	$ \pm$ 	0.19	    &  		\nodata		      \\
2451893.84	 & 	1.678	 &  	19.39	$ \pm$	0.10	    &  	20.25	$ \pm $	0.28	      \\
2451896.85	 & 	1.721	 &  	19.21	$ \pm $	0.07	    &  	19.95	$ \pm $	0.22	      \\
2451899.86	 & 	1.760	 &  	19.37	$ \pm $	0.10	    &  		\nodata		      \\
2451901.84	 & 	1.782	 &  	19.40	$ \pm $	0.06	    &  	20.08	$ \pm $	0.14	      \\
2451902.86	 & 	1.793	 &  	19.25	$ \pm $	0.08	    &  		\nodata		      \\
2451903.81	 & 	1.803	 &  	19.45	$ \pm $	0.06	    &  	20.10	$ \pm $	0.08	      \\
2451904.86	 & 	1.813	 &  	19.20	$ \pm $	0.11	    &  		\nodata		      \\
2451906.85	 & 	1.831	 &  	19.45	$ \pm $	0.04	    &  	19.89	$ \pm $	0.09	      \\
2451907.84	 & 	1.839	 &  	19.45	$ \pm $	0.04	    &  	20.09	$ \pm $	0.08	      \\
2451908.86	 & 	1.846	 &  	19.41	$ \pm $	0.05	    &  		\nodata		      \\
2451909.85	 & 	1.853	 &  	19.49	$ \pm $	0.05	    &  		\nodata		      \\
2451912.80	 & 	1.871	 &  	19.38	$ \pm $	0.05	    &  	19.97	$ \pm $	0.09	      \\
2451914.82	 & 	1.879	 &  	19.41	$ \pm $	0.03	    &  		\nodata		      \\
2451915.82	 & 	1.883	 &  	19.33	$ \pm $	0.04	    &  	20.14	$ \pm $	0.08	      \\
2451917.86	 & 	1.889	 &  	19.38	$ \pm $	0.06	    &  	19.98	$ \pm $	0.06	      \\
2451918.84	 & 	1.890	 &  	19.47	$ \pm $	0.06	    &  	20.10	$ \pm $	0.13	      \\
2451919.81	 & 	1.892	 &  	19.31	$ \pm $	0.09	    &  	19.77	$ \pm $	0.14	      \\
2451920.79	 & 	1.893	 &  		\nodata		    &  	19.93	$ \pm $	0.14	      \\
2451921.85	 & 	1.893	 &  	19.45	$ \pm $	0.06	    &  	20.08	$ \pm $	0.15	      \\
2451922.78	 & 	1.893	 &  	19.33	$ \pm $	0.09	    &  	20.27	$ \pm $	0.26	      \\
2451923.81	 & 	1.892	 &  	19.40	$ \pm $	0.07	    &  		\nodata		      \\
2451927.75	 & 	1.883	 &  	19.35	$ \pm $	0.09	    &  	19.77	$ \pm $	0.10	      \\
2451928.80	 & 	1.879	 &  	19.38	$ \pm $	0.03	    &  	20.03	$ \pm $	0.07	      \\
2451929.78	 & 	1.874	 &  	19.37	$ \pm $	0.04	    &  	20.03	$ \pm $	0.08	      \\
2451930.80	 & 	1.869	 &  	19.41	$ \pm $	0.05	    &  		\nodata		      \\
2451931.78	 & 	1.864	 &  	19.48	$ \pm $	0.04	    &  	20.18	$ \pm $	0.08	      \\
2451932.77	 & 	1.858	 &  	19.30	$ \pm $	0.03	    &  	19.99	$ \pm $	0.06	      \\
2451933.80	 & 	1.851	 &  	19.38	$ \pm $	0.03	    &  	19.96	$ \pm $	0.06	      \\
2451936.77	 & 	1.828	 &  	19.33	$ \pm $	0.03	    &  	19.94	$ \pm $	0.06	      \\
2451937.82	 & 	1.819	 &  	19.34	$ \pm $	0.04	    &  	19.88	$ \pm $	0.06	      \\
2451938.82	 & 	1.810	 &  	19.32	$ \pm $	0.03	    &  	19.95	$ \pm $	0.06	      \\
2451939.82	 & 	1.800	 &  	19.38	$ \pm $	0.04	    &  	19.91	$ \pm $	0.05	      \\
2451940.80	 & 	1.790	 &  	19.30	$ \pm $	0.05	    &  	20.07	$ \pm $	0.09	      \\
2451941.78	 & 	1.779	 &  	19.34	$ \pm $	0.03	    &  	20.00	$ \pm $	0.06	      \\
2451942.73	 & 	1.767	 &  	19.30	$ \pm $	0.04	    &  	20.00	$ \pm $	0.08	      \\
2451954.71	 & 	1.587	 &  	19.38	$ \pm $	0.05	    &  		\nodata		      \\
2451955.74	 & 	1.568	 &  	19.31	$ \pm $	0.03	    &  		\nodata		      \\
2451957.78	 & 	1.530	 &  	19.25	$ \pm $	0.03	    &  		\nodata		      \\
2451960.78	 & 	1.470	 &  	19.33	$ \pm $	0.03	    &  		\nodata		      \\
2451966.80	 & 	1.336	 &  	19.25	$ \pm $	0.03	    &  		\nodata		      \\
2451971.80	 & 	1.213	 &  	19.28	$ \pm $	0.03	    &  		\nodata		      \\
2451974.77	 & 	1.136	 &  	19.24	$ \pm $	0.03	    &  		\nodata		      \\
2451981.73	 & 	0.947	 &  	19.21	$ \pm $	0.04	    &  		\nodata		      \\
2451984.72	 & 	0.862	 &  	19.18	$ \pm $	0.03	    &  		\nodata		      \\
2451987.83	 & 	0.775	 &  	19.17	$ \pm $	0.03	    &  		\nodata		      \\
2451991.80	 & 	0.655	 &  	19.16	$ \pm $	0.03	    &  		\nodata		      \\
2451994.75	 & 	0.571	 &  	19.18	$ \pm $	0.03	    &  		\nodata		      \\
2451997.73	 & 	0.492	 &  	19.17	$ \pm $	0.03	    &  		\nodata		      \\
2452001.71	 & 	0.392	 &  	19.14	$ \pm $	0.02	    &  		\nodata		      \\
2452004.74	 & 	0.329	 &  	19.15	$ \pm $	0.04	    &  		\nodata		      \\
2452008.75	 & 	0.284	 &  	19.15	$ \pm $	0.10	    &  		\nodata		      \\
2452011.61	 & 	0.287	 &  	19.11	$ \pm $	0.03	    &  		\nodata		      \\
2452026.58	 & 	0.619	 &  	19.09	$ \pm $	0.03	    &  		\nodata		      \\
2452030.73	 & 	0.745	 &  	19.19	$ \pm $	0.04	    &  	19.80	 $\pm$ 	0.06	      \\
2452031.59	 & 	0.765	 &  	19.21	$ \pm $	0.04	    &  		\nodata		      \\
2452040.64	 & 	1.024	 &  	19.25	$ \pm $	0.03	    &  		\nodata		      \\
2452045.77	 & 	1.164	 &  	19.37	$ \pm $	0.07	    &  		\nodata		      \\
2452046.61	 & 	1.186	 &  	19.21	$ \pm $	0.03	    &  		\nodata		      \\
2452048.67	 & 	1.238	 &  	19.27	$ \pm $	0.03	    &  		\nodata		      \\
2452051.60	 & 	1.312	 &  	19.26	$ \pm $	0.03	    &  		\nodata		      \\
2452053.52	 & 	1.357	 &  	19.26	$ \pm $	0.03	    &  		\nodata		      \\
2452054.56	 & 	1.382	 &  	19.26	$ \pm $	0.04	    &  		\nodata		      \\
2452055.57	 & 	1.405	 &  	19.31	$ \pm $	0.03	    &  		\nodata		      \\
2452057.57	 & 	1.451	 &  	19.28	$ \pm $	0.03	    &  		\nodata		      \\
2452066.59	 & 	1.633	 &  	19.23	$ \pm $	0.05	    &  		\nodata		      \\
2452067.58	 & 	1.651	 &  	19.34	$ \pm $	0.04	    &  		\nodata		      \\
2452068.53	 & 	1.667	 &  	19.30	$ \pm $	0.03	    &  		\nodata		      \\
2452069.51	 & 	1.683	 &  	19.41	$ \pm $	0.03	    &  		\nodata		      \\
2452070.51	 & 	1.700	 &  	19.34	$ \pm $	0.05	    &  		\nodata		      \\
2452075.51	 & 	1.776	 &  	19.37	$ \pm $	0.03	    &  		\nodata		      \\
2452094.51	 & 	1.946	 &  	19.46	$ \pm $	0.06	    &  		\nodata		      \\
2452096.60	 & 	1.952	 &  	19.38	$ \pm $	0.09	    &  		\nodata		      \\
2452101.51	 & 	1.960	 &  	19.34	$ \pm $	0.06	    &  		\nodata		      \\
2452102.49	 & 	1.960	 &  	19.44	$ \pm $	0.03	    &  		\nodata		      \\
2452105.50	 & 	1.956	 &  	19.44	$ \pm $	0.04	    &  		\nodata		      \\
2452112.49	 & 	1.930	 &  	19.49	$ \pm $	0.09	    &  		\nodata		      \\
2452115.52	 & 	1.910	 &  	19.32	$ \pm $	0.05	    &  		\nodata		      \\ 
\enddata

\end{deluxetable}

\clearpage

\begin{deluxetable}{ccc}
\tabletypesize{\scriptsize}
\tablecaption{Five-Day Binned Light Curve for 2000 $EB_{173}$. \label{tbl-2}}
\tablewidth{0pt}
\tablehead{
\colhead{$\langle HJD \rangle$} & \colhead{$< \alpha >$}   & \colhead{$R_{OPP}$}
}
\startdata
2451892.3	&	1.66	&  $	19.25	 \pm 	0.09	 $     \\
2451898.4	&	1.74	&  $	19.16	 \pm 	0.06	 $     \\
2451903.3	&	1.80	&  $	19.27	 \pm 	0.04	 $     \\
2451908.3	&	1.84	&  $	19.35	 \pm 	0.02	 $     \\
2451913.8	&	1.88	&  $	19.31	 \pm 	0.03	 $     \\
2451918.1	&	1.89	&  $	19.30	 \pm 	0.03	 $     \\
2451922.3	&	1.89	&  $	19.33	 \pm 	0.04	 $     \\
2451928.8	&	1.88	&  $	19.31	 \pm 	0.03	 $     \\
2451932.3	&	1.86	&  $	19.32	 \pm 	0.02	 $     \\
2451938.3	&	1.81	&  $	19.28	 \pm 	0.02	 $     \\
2451941.8	&	1.78	&  $	19.27	 \pm 	0.02	 $     \\
2451954.7	&	1.59	&  $	19.34	 \pm 	0.05	 $     \\
2451956.8	&	1.55	&  $	19.25	 \pm 	0.03	 $     \\
2451960.8	&	1.47	&  $	19.30	 \pm 	0.03	 $     \\
2451966.8	&	1.34	&  $	19.23	 \pm 	0.03	 $     \\
2451973.3	&	1.17	&  $	19.25	 \pm 	0.02	 $     \\
2451983.2	&	0.90	&  $	19.18	 \pm 	0.03	 $     \\
2451987.8	&	0.78	&  $	19.16	 \pm 	0.03	 $     \\
2451993.3	&	0.61	&  $	19.17	 \pm 	0.02	 $     \\
2451997.7	&	0.49	&  $	19.16	 \pm 	0.03	 $     \\
2452003.2	&	0.36	&  $	19.14	 \pm 	0.02	 $     \\
2452008.8	&	0.28	&  $	19.15	 \pm 	0.10	 $     \\
2452011.6	&	0.29	&  $	19.11	 \pm 	0.03	 $     \\
2452026.6	&	0.62	&  $	19.09	 \pm 	0.03	 $     \\
2452031.2	&	0.76	&  $	19.20	 \pm 	0.03	 $     \\
2452040.6	&	1.02	&  $	19.24	 \pm 	0.03	 $     \\
2452047.0	&	1.20	&  $	19.24	 \pm 	0.02	 $     \\
2452053.2	&	1.35	&  $	19.24	 \pm 	0.02	 $     \\
2452056.6	&	1.43	&  $	19.27	 \pm 	0.02	 $     \\
2452068.1	&	1.66	&  $	19.32	 \pm 	0.02	 $     \\
2452070.5	&	1.70	&  $	19.30	 \pm 	0.05	 $     \\
2452075.5	&	1.78	&  $	19.33	 \pm 	0.03	 $     \\
2452094.5	&	1.95	&  $	19.40	 \pm 	0.06	 $     \\
2452096.6	&	1.95	&  $	19.32	 \pm 	0.09	 $     \\
2452102.0	&	1.96	&  $	19.35	 \pm 	0.03	 $     \\
2452105.5	&	1.96	&  $	19.37	 \pm 	0.04	 $     \\
2452114.0	&	1.92	&  $	19.28	 \pm 	0.04	 $     \\
\enddata

\end{deluxetable}

\clearpage

\begin{deluxetable}{cccl}
\tabletypesize{\scriptsize}
\tablecaption{\vr ~ Color for 2000 $EB_{173}$. \label{tbl-3}}
\tablewidth{0pt}
\tablehead{
\colhead{$< \alpha >$}   &	\colhead{$\vr$} & \colhead{$V_{OPP}$}   & \colhead{Reference}
}
\startdata
0.75	&   $	0.62	 \pm 	0.07	 $   &	19.80	&	This paper	   \\
0.81	&   $	0.34	 \pm 	0.18	 $   &	19.52	&	\citet{fer01}	   \\
0.81	&   $	0.73	 \pm 	0.16	 $   &	19.91	&	\citet{fer01}	   \\
0.92	&   $	0.61	 \pm 	0.15	 $   &	19.81	&	\citet{fer01}	   \\
0.95	&   $	0.40	 \pm 	0.18	 $   &	19.60	&	\citet{fer01}	   \\
1.01	&   $	0.76	 \pm 	0.16	 $   &	19.97	&	\citet{fer01}	   \\
1.14	&   $	0.67	 \pm 	0.18	 $   &	19.90	&	\citet{fer01}	   \\
1.19	&   $	0.54	 \pm 	0.15	 $   &	19.77	&	\citet{fer01}	   \\
1.59	&   $	0.44	 \pm 	0.16	 $   &	19.72	&	\citet{fer01}	   \\
1.60	&   $	0.69	 \pm 	0.18	 $   &	19.97	&	\citet{fer01}	   \\
1.73	&   $	0.60	 \pm 	0.10	 $   &	19.90	&	\citet{fer01}	   \\
1.78	&   $	0.63	 \pm 	0.03	 $   &	19.94	&	This paper	   \\
1.80	&   $	0.55	 \pm 	0.05	 $   &	19.86	&	\citet{fer01}	   \\
1.82	&   $	0.63	 \pm 	0.02	 $   &	19.94	&	This paper	   \\
1.93	&   $	0.65	 \pm 	0.03	 $   &	19.97	&	\citet{jel01}	   \\
\enddata

\end{deluxetable}

\clearpage

\begin{figure}
\plotone{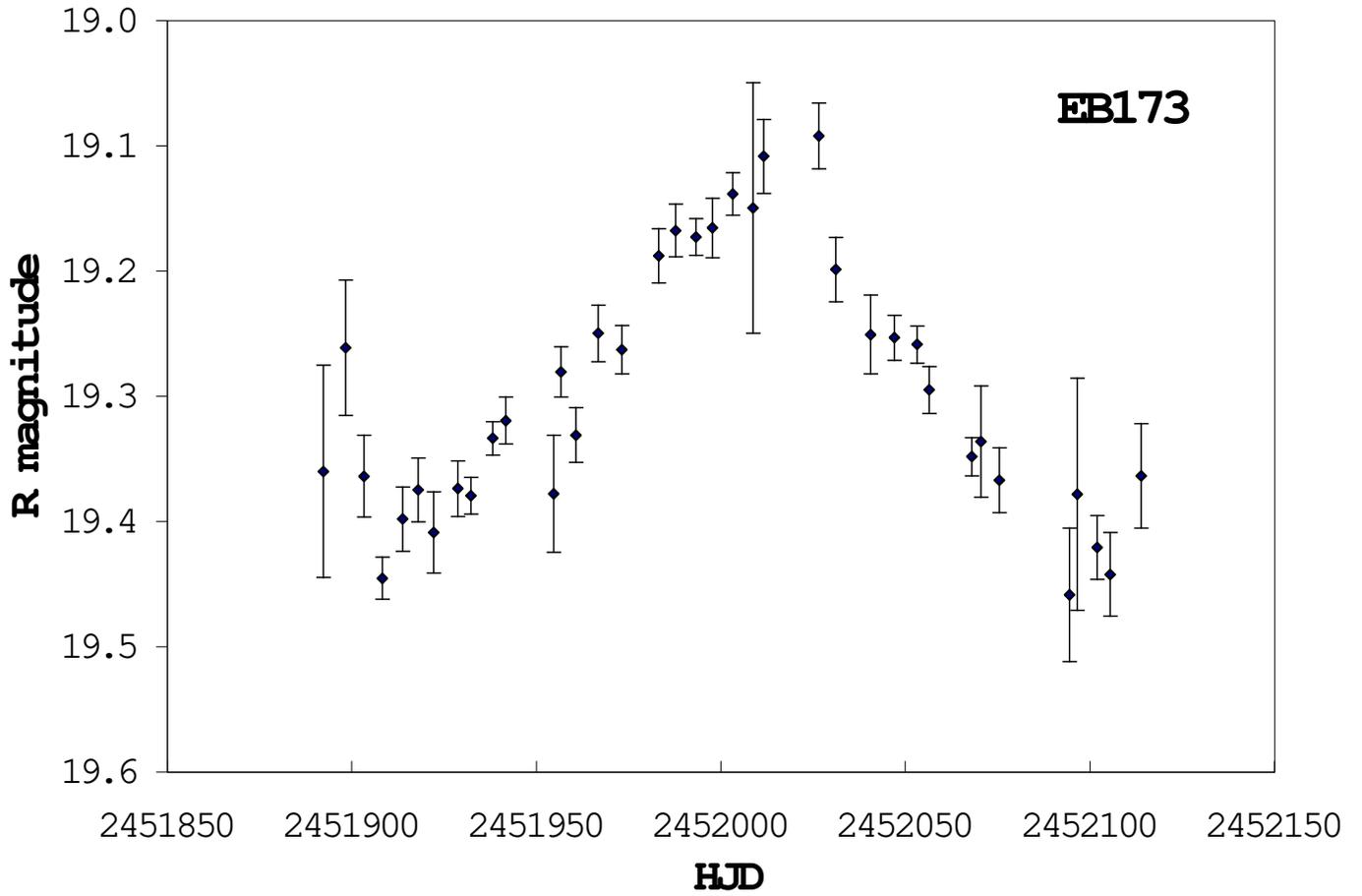}
\vspace{2.in}
\caption{Light curve for 2000 $EB_{173}$.  This light curve plots the 
five-day binned R-band magnitudes of 2000 $EB_{173}$ 
corrected to opposition distance, as taken from Table 2. \label{fig1}}
\end{figure}

\clearpage 

\begin{figure}
\plotone{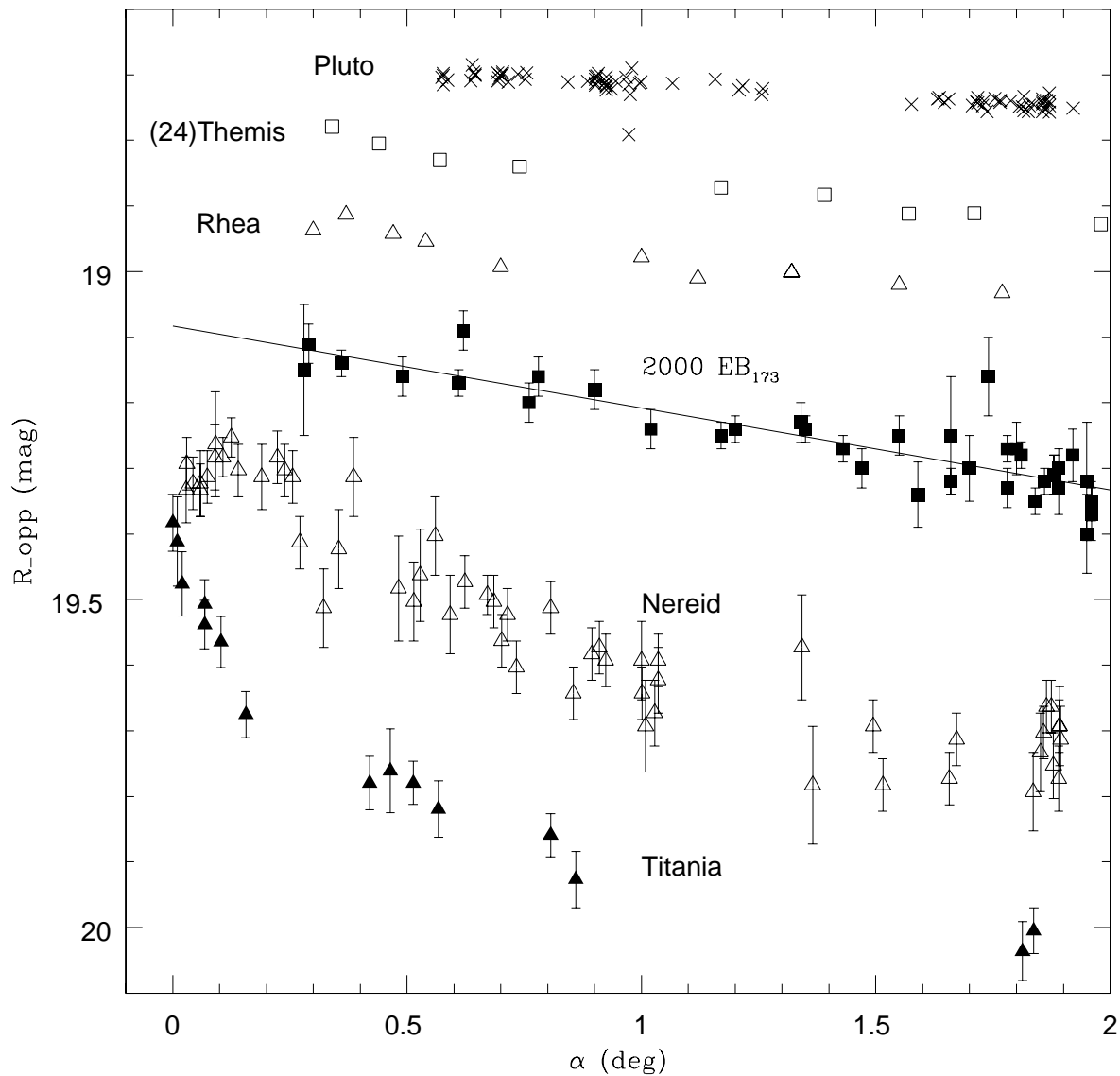}
\caption{Phase curve for 2000 $EB_{173}$ and other bodies.  From top to bottom, the phase 
functions for Pluto (Tholen \& Tedesco 1994), (24) 
Themis (Harris et al. 1989), Rhea (Domingue, Lockwood, \& Thompson 1995),  2000 $EB_{173}$ 
(this paper), Nereid (Schaefer \& Tourtellotte 2001), and Titania (Buratti, Gibson, \& Mosher 
1992)  The phase function for 2000 $EB_{173}$ is well fit by $R_{OPP}=19.083+0.125 \alpha$ 
for $0.284 \degr < \alpha < 1.910 \degr$, as shown by the straight line. For all other 
phase functions, arbitrary vertical offsets have been added.\label{fig2}}
\end{figure}

\end{document}